\documentclass[twocolumn,showpacs,preprintnumbers,amsmath,amssymb]{revtex4}

\usepackage{graphicx}
\usepackage{dcolumn}
\usepackage{bm}
\graphicspath{{pictures/}} 

\begin{document}

\preprint{APS/123-QED}

\title{Late growth stages and post-growth diffusion in organic epitaxy: PTCDA on Ag(111)} 

\author{B. Krause}
\altaffiliation[New address: ]{European Synchrotron Radiation Facility, BP 220,  F-38043 Grenoble Cedex, France}

\author{A.~C.~D\"urr}%
\author{F. Schreiber}%
\altaffiliation[Author for correspondence; new address: ]{Physical and Theoretical Chemistry Laboratory, Oxford University, South Parks Road, Oxford, OX1 3QZ, United Kingdom;
                email address: frank.schreiber@chem.ox.ac.uk}
\author{H. Dosch}%
\affiliation{
         Max-Planck-Institut f\"ur Metallforschung, Heisenbergstr.3,
         70569 Stuttgart
   and
   Institut f\"ur Theoretische und Angewandte Physik,
         Universit\"at Stuttgart, Pfaffenwaldring 57, 70550 Stuttgart,
         Germany
    }

\author{O.~H.~Seeck}
\affiliation{IFF, Forschungszentrum J\"ulich GmbH, D-52425 J\"ulich, Germany}%

\date{\today}

\begin{abstract}

The late growth stages and the post-growth diffusion of crystalline 
organic thin films have been investigated for 
3,4,9,10-perylenetetracarboxylic dianhydride
(PTCDA)
on Ag(111), 
a model system in organic epitaxy. 
In situ x-ray measurements at the anti-Bragg point during the growth show 
intensity oscillations followed by a time-independent intensity 
which is independent of the growth temperature. 
At $T \gtrsim $ 350 K, the intensity increases after growth up to a 
temperature-dependent saturation value due to a post-growth diffusion process. 
The time-independent intensity and the subsequent intensity recovery 
have been reproduced by models based on the morphology change as a function 
of the growth temperature. 
The morphology found after the post-growth diffusion processes has been 
studied by specular rod measurements.  

\end{abstract}

\maketitle

\section{Introduction}
\label{Introduction}
Crystalline small-molecule organic semiconductors such as perylene and phthalocyanine derivatives 
show interesting  electrical and optical properties, which
can be exploited in electronic and optoelectronic devices 
based on thin films deposited by organic molecular beam epitaxy (OMBE). 
However, the intrinsic properties of the organic crystals are closely related 
to the molecular and crystalline anisotropy, and 
the crystalline quality of the organic thin film has a strong impact 
on the resulting device performance. 
In order to control the properties of organic thin films, 
the understanding of the basic growth processes is necessary.

While the structure and epitaxy has been thoroughly studied
for several model systems, such as PTCDA 
\cite{Forrest97,Hooks01,Glockler98,KrausePRB02,Seidel01,Stohr02,SchmitzHubsch99,Fenter97},
on Ag(111) 
the growth dynamics and evolution of the resulting morphology
have been less systematically investigated.
Importantly, even for a system with well-defined monolayer
epitaxy, the later stages of the growth (after a certain
threshold thickness) can, of course, exhibit islanding
(Stranski-Krastanov (SK) growth) and a very rough resulting morphology.
It was recently found that PTCDA on Ag(111), one of the
structurally best characterized OMBE systems,
exhibits a relatively smooth morphology (kinetically limited)
at low temperatures ($T \lesssim 350$~K),
but that the growth behavior undergoes a transition to a
very clear SK scenario at high temperatures ($T \gtrsim 350$~K).
In this paper, we present a real-time study of the later stages
of the growth (islanding process) and the post-growth
diffusion processes, which have a profound impact on the
resulting morphology. We attempt a comprehensive discussion of the
effects related to the temperature dependent transition
of the morphology.

\begin{figure}[tbp!]
\centering
\includegraphics[width=0.85\columnwidth]{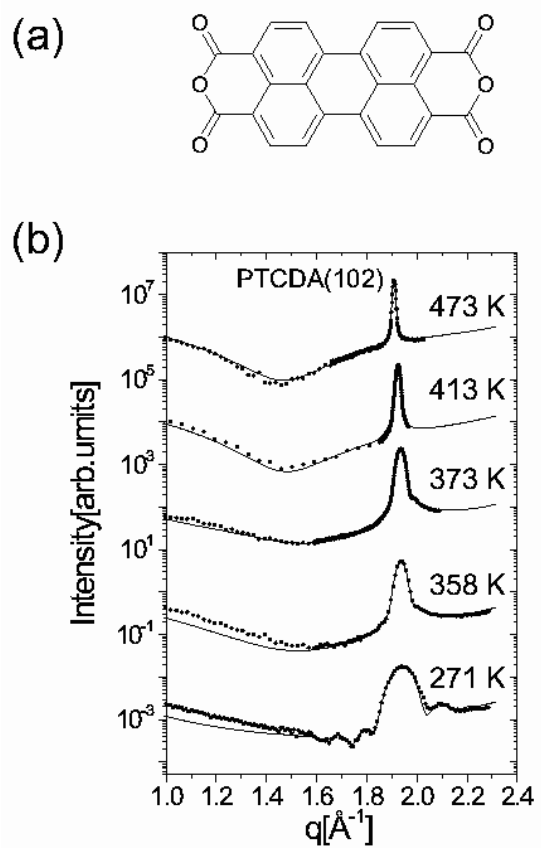}
\caption{(a) 3,4,9,10-perylenetetracarboxylic dianhydride (PTCDA). 
(b) Measured (circles) and calculated (lines) specular rod intensity as a 
function of the growth temperature $T$ for samples with 
$F=0.8-2$~\AA/min and $d=50$~\AA.}
\label{Fig1}
\end{figure}

PTCDA (C$_{24}$O$_6$H$_8$, see Fig.\ \ref{Fig1} (a)) 
is an archetypal system for OMBE \cite{Forrest97}. 
It is a flat molecule which crystallizes in two monoclinic bulk polymorphs, 
the $\alpha$ and the $\beta$ phase \cite{Lovinger84,Kaplan,Mobus92,Ogawa99}. 
In the following, we will always refer to the $\alpha$ bulk notation. 
For both polymorphs, the 2 molecules of the unit cell order in a herringbone structure 
with the molecular plane parallel to the PTCDA(102) plane. 
At room temperature, the stacking distance between subsequent molecular sheets is 
$d^\alpha_F=3.22$~\AA \ and $d^\beta_F=3.25$~\AA \ in PTCDA(102) direction. 

On Ag(111), PTCDA forms well-ordered epitaxial films \cite{Glockler98,KrausePRB02}. 
It has been found that the binding between the commensurate first monolayer 
and the substrate is much stronger than the binding between the subsequent 
layers \cite{Taborski95}. 
The structure and the morphology of PTCDA films vary significantly with the 
growth temperature, $T$ \cite{KrausePRB02,Chkoda03}. 
For $T \lesssim 350$~K, the films are strained and have a relatively homogenous thickness. 
For $T \gtrsim 350$~K, the films consist of relaxed islands on top of two wetting layers. 
For a variety of growth conditions, a coexistence of both bulk polymorphs 
has been observed. 

A detailed study of the epitaxy and the post-growth morphology of the PTCDA films 
studied here has been published elsewhere 
\cite{Krause01,KrausePRB02,Krausechemphys03,Krauseosc}. 
In this paper, we focus on the results of specular intensity measurements 
(with the momentum transfer ${\bf q}=q_z {\bf e}_z$ perpendicular to the sample surface) 
during and directly after the growth of thin PTCDA films. 
This type of x-ray measurement is very suitable for studying the evolution 
of the out-of-plane structure and morphology of organic thin films in real time. 
It contains the laterally averaged information about the thin film structure 
in $z$-direction. 

Two types of experiments are reported in the following: 
measurements of a section of the specular rod, 
i.e.\ intensity measurements as a function of $q_z$, after the growth, 
and time-dependent intensity measurements at one specific point of the specular rod 
during and after the deposition of the PTCDA films. 
For this, the anti-Bragg point of the film with $q_z^\ast=\pi/d_F$ has been chosen 
(see below).
 
The first type of experiment gives insight into the crystalline structure of the films. 
It will be shown how the growth temperature dependence of the morphology influences 
the specular intensity. 
The second type of experiment gives time-dependent information about the 
material distribution between the different layers. 
It contains information about the growth process and the post-growth diffusion.
The initial stages of the growth are described in detail elsewhere \cite{Krauseosc}. 
Here, the later stages of the growth (after the growth of the wetting layer) 
and the post-growth diffusion are discussed.

The paper is organized as follows. In Sec.\ \ref{Experimental} the details 
about the sample preparation are given. 
The experiments are described in Sec.\ \ref{Results}. 
In Sec.\ \ref{Discussion}, the results are discussed, 
and theoretical models for the observations are developed. 
The main results and conclusions of the paper are summarized 
in Sec.\ \ref{Summary and Conclusions}.

%
\section{Experimental}
\label{Experimental}

The organic thin films have been deposited by molecular beam epitaxy in a portable 
UHV chamber (base pressure $10^{-10}$~mbar) equipped for x-ray experiments \cite{Ritley01}. 
A Ag(111) single crystal with an out-of-plane mosaicity of 
$0.13 \pm 0.02^{\circ}$ has been used as a substrate. 
Prior to deposition, the substrate has been 
cleaned by repeated cycles of  sputtering 
(30~min at room temperature, Ar$^+$ energy 500--600~eV, 
Ar pressure $\approx 6 \times 10^{-5}$~mbar) and annealing at $\approx 700-900$~K 
for at least 15~min. 
The purified and thoroughly outgassed PTCDA has been evaporated from a Knudsen cell 
kept at approximately 575~K. 

Two sets of samples with the average film thickness $d=50$~\AA \ 
and growth temperatures $T$ in the range of 197 -- 473~K have been studied, 
one with a flux (growth rate) $F=0.8 - 2.0$~\AA/min (low $F$), 
the other with $F=12-14$~\AA/min (high $F$). 

The experiments have been performed at the Hamburger Synchrotronstrahlungslabor (HASYLAB) 
at the beamline W1 with the wavelength $\lambda=1.237$~\AA.

\section{Results}
\label{Results}
\subsection{Specular rod measurements}
\label{Specular rod measurements}
Fig.\ \ref{Fig1} shows examples of the specular rod measurements including the 
PTCDA(102) reflection for samples grown at low $F$. 
The measurements have been performed at the respective growth temperature of the samples. 
Independent of $T$, the PTCDA(102) plane is parallel to Ag(111). 
For $T < 358$~K, 1--3 Laue oscillations around PTCDA(102) have been observed. 
In this temperature range, the width of the PTCDA(102) peak corresponds 
to the average film thickness and is independent of $T$. 
For $T \geq 358$~K, the peak width of the Bragg peak decreases with increasing $T$, 
and the Laue oscillations reduce to two shoulders on the Bragg peak or disappear 
completely. 
With increasing $T$, the rod develops a pronounced minimum at $q_z\approx 1.4$~\AA$^{-1}$. 
While the rod measurements show a strong temperature-dependence, 
the rocking scans through PTCDA(102) are independent of $T$ and show the same peak width 
of $0.13 \pm 0.02^\circ$ corresponding to the mosaicity of the substrate. 
A detailed discussion of the modelling of the data is presented in 
Sec.\ \ref{Modelling of the rod measurements}.

The measurements for the high $F$ samples (not shown here) show a similar behavior, 
but the changes in the rod measurements and the peak shape are shifted to 
higher temperatures by about 15~K. The Laue oscillations and the peak width expected 
for the average film thickness are still observed for $T \leq 358$~K, 
where the oscillations are already damped out for the low $F$ samples.
\subsection{In situ experiments}
\label{In situ experiments}
\begin{figure}[tbp!]
\centering
\includegraphics[width=\columnwidth]{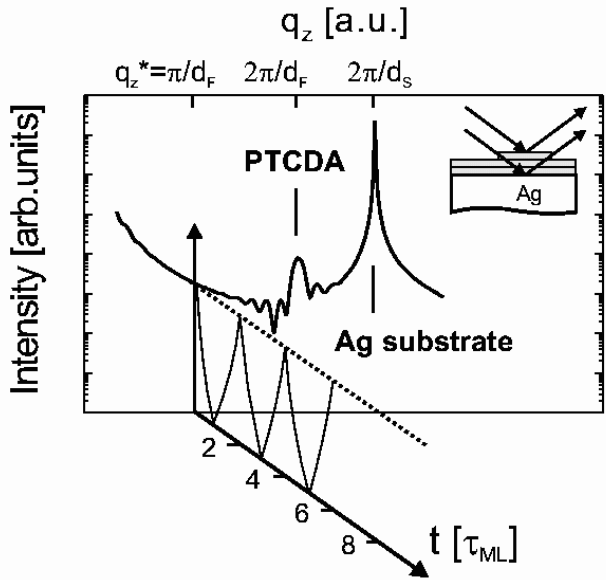}
\caption{Simulation of the specular rod of a thin PTCDA film on Ag(111). At the anti-Bragg point of the PTCDA film ($q_z^\ast=\pi/d_F$), the scattering of subsequent layers interferes destructively (see inset). For layer-by-layer growth, the intensity at the anti-Bragg point oscillates as a function of the deposition time.}
\label{Fig2b}
\end{figure}
In order to follow the growth of thin PTCDA films, 
the x-ray diffraction intensity from the molecular layers has been measured in real-time. 
In kinematic approximation the specular x-ray scattering
is the sum of the scattering contributions from the film and the substrate,
\begin{eqnarray}
\label{Eq1}
S(q_z,t) &=& \bigg[ f_S(q_z) \left( \frac{1}{1-e^{-i q_z d_S}}+e^{i q_z d_{S1}}\right)  e^{-i q_z(d_0+d_{S1})} \nonumber \\
&+& f_F(q_z)  \sum_{k=1}^\infty \theta_k(t) e^{i(k-1) q_z d_F} \bigg ] e^{-\frac{\Delta q_z^2 \sigma_S^2}{2}}. 
 \end{eqnarray}
Here, $\Delta q_z$ is the difference between $q_z$ and the nearest specular Bragg peak of the substrate with $q_z=l 2 \pi/d_S$. $d_F$ and $d_S$ are the  lattice spacings of the film and the substrate, $f_F$ and $f_S$ are the corresponding form factors, $d_{s1}$ is the lattice spacing of the topmost layer of the substrate, $d_0$  is the distance between the substrate and the first layer of the film, $\theta_n$ is the fractional coverage of the $k$th layer within the film, and $\sigma_S$ is the Gaussian roughness of the substrate. At the anti-Bragg point of the PTCDA film ($q_z^\ast=\pi/d_F$) the contribution of the substrate (the second term in Eq.\ \ref{Eq1}) equals $f_F\sum (-1)^{(k-1)} \theta_k(t)$, which can be written as $f_F \Delta \theta$, where $\Delta \theta$ is defined as
\begin{equation}
\label{Eq2}
\Delta \theta(t)= \sum_{m}\theta_{2m+1}(t) - \sum_{m}\theta_{2m}(t)=\theta_{odd}(t)-\theta_{even}(t). 
\end{equation}
This is the difference between the material (in fractions of monolayers) deposited in odd layers (i.e.\ the coverage of layer 1 $+$ the coverage of layer 3 $+$ ...) and the material deposited in even layers. This simplification is possible since at the anti-Bragg point the scattering amplitude of all even and all odd layers has the same phase. It has been assumed here that the lattice parameter of the film is constant across the entire film thickness. This assumption is supported by the specular rod measurements.

The scattered intensity at the anti-Bragg point, normalized to the scattering of the substrate, is
\begin{equation}
\label{Eq3}
I(q_z^\ast,t)=1+A \Delta \theta + B (\Delta \theta)^2
\end{equation}
with 
\begin{eqnarray}
\label{Eq4}
A &=& 2 Re \left(f_F^\ast f_s \left( \frac{1}{1-e^{-iq_z^\ast d_S}}+e^{iq_z^\ast d_{S1}}  \right) \; e^{-iq_z^\ast \left(d_0+d_{S1} \right)} \right) \nonumber \\
& & \times \frac{4 \sin \left(q_z^\ast d_F/2 \right)}{\left| f_s \right|^2} \nonumber \\
B &=& \frac{ \left|f_F \right|^2 4 \sin \left(q_z^\ast d_F/2 \right)
 }{ \left|f_S \right| ^2} 
\end{eqnarray}

The parameters $A$ and $B$ depend on the materials involved. For PTCDA on Ag(111), inserting the atomic form factors and the lattice parameters determined from the rod measurements (see Sec.\ \ref{Specular rod measurements}) leads to the values $A=-0.73$ and $B=0.06$.

In the case of layer-by layer growth, characteristic intensity oscillations with the minimum intensity $I_{min}=1+A+B$ and the maximum intensity $I_{max}=1$  are observed as a function of the deposition time \cite{Vlieg88,Dai99,Krauseosc,Kaganer03} (see Fig.\ \ref{Fig2b}).\\

\begin{figure}[tbp!]
\centering
\includegraphics[width=\columnwidth]{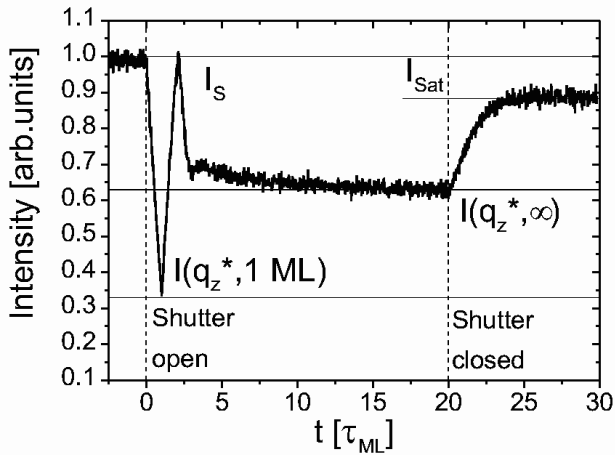}
\caption{Typical intensity measurement at the anti-Bragg point during the deposition 
of a film with $T=393$~K and $F \approx 1$~\AA/min. }
\label{Fig3}
\end{figure}
Fig.\ \ref{Fig3} shows a typical intensity measurement during the 
deposition of a PTCDA film, 
normalized to the intensity scattered by the substrate measured before deposition. 
The time is given in units  of the monolayer deposition time, $\tau_{ML}$, 
of the first few layers as derived by the observed intensity oscillations. 
The temperature-dependence of the intensity oscillations during the deposition of 
the initial monolayers has been described in detail elsewhere \cite{Krauseosc}. 
In the following we briefly summarize the most important observations, and then 
focus on the analysis of the later stages of the growth 
and the post-growth changes.

Before opening the shutter of the evaporation cell, the constant intensity, $I_S$, scattered by the substrate is measured. During deposition, the intensity oscillates between the maximum value $I(q_z^\ast, 2\tau_{ML}) \lesssim 1 \pm 0.02$ and the minimum value $I(q_z^\ast, \tau_{ML}) \gtrsim 0.33 \pm 0.02$ for $t \lesssim 3 \tau_{ML}$. $\tau_{ML}$ is the deposition time for 1~ML. For $T \lesssim 358$~K the intensity oscillations for $t \lesssim 3 \tau_{ML}$ are damped, for $T > 358$~K they are undamped and show no temperature dependence. Interestingly, during further deposition, the intensity approximates the asymptotic value $I(q_z^\ast, \infty) = 0.65 \pm 0.02$, independent of $T$.

\begin{figure}[tbp!]
\centering
\includegraphics[width=0.85\columnwidth]{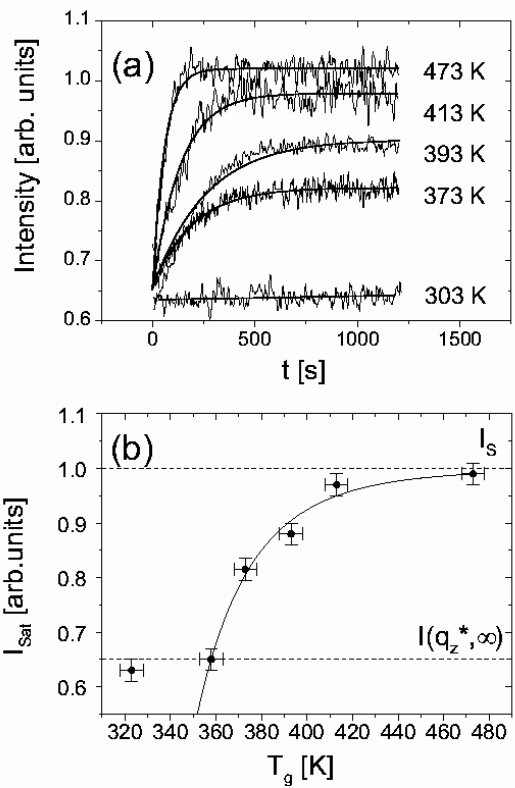}
\caption{(a)  Intensity at the anti-Bragg point after the deposition 
of a film as a function of $T$, measured for several samples with  
$d \approx 50$~\AA \ and $F = 0.8$ -- $2$~\AA/min. 
The shutter of the evaporation cell has been closed at $t=0$. 
(b) Saturation intensity as a function of $T$.}
\label{Fig4}
\end{figure}

After the shutter has been closed, the intensity increases with $t$ up to a saturation value, $I_{Sat} > I(q_z^\ast, \infty)$, indicative of post-growth structural changes. Fig.\ \ref{Fig4} (a) shows this intensity recovery as a function of the growth temperature. For $T \leq  358$~K, $I(q_z^\ast,t) = I(q_z^\ast, \infty)=0.65 \pm 0.02$ remains constant after deposition, while for $T > 358$~K, the intensity increases as a function of the time after interrupting the deposition. The saturation intensity shown in Fig.\ \ref{Fig4} (b) increases with $T$ up to $ I_{Sat}=1 \pm 0.03$ , while the characteristic time, $\tau_{Sat}$, for the saturation process decreases with $T$ (see Sec.\ \ref{Discussion} for a quantitative analysis).

\section{Discussion}
\label{Discussion}
For the PTCDA films discussed here, a morphology change from a smooth or 
facetted film to islands on a wetting layer has been observed as a function of 
$T$ \cite{KrausePRB02,Krauseosc}. 
The transition to the island morphology has been found around 
$T \approx 350$~K which is comparable to the temperature where the changes 
in the rod measurements and in the {\sl in situ} intensity measurements 
have been observed. 
This suggests that the observations reported here are related to the morphology change.

\subsection{Rod measurements}
\label{Discussion of the rod measurements}
Qualitatively, the observed features of the rod measurements correspond 
to the following scenario. 
The PTCDA films grown at low temperatures (for the low $F$ samples, this 
corresponds to $T <358$~K) are relatively smooth as indicated by the Laue oscillations. 
The films cover essentially the entire substrate surface since the 
width of the Bragg peak corresponds to the average film thickness. 
For higher temperatures, the Laue oscillations vanish indicating an 
increasing roughness of the film. 
The decreasing width of the Bragg peak corresponds to high islands, 
and the intensity minimum at $q_z\approx 1.4$~\AA$^{-1}$ indicates a 2~ML 
thick wetting layer between the islands. 
The observations agree with the morphology observed by AFM \cite{KrausePRB02}, 
implying that the entire PTCDA film is ordered. 

For all films, only a single PTCDA out-of-plane Bragg peak 
around $q_z\approx 1.95$~\AA$^{-1}$ has been observed, 
even if the in-plane measurements have confirmed the coexistence of the 
$\alpha$ and the $\beta$ phase \cite{KrausePRB02}. 
The expected splitting between the (102) Bragg reflections of both polymophs 
is $\Delta Q=2\pi/d_F^\alpha-2\pi/d_F^\beta   \approx 0.02$~\AA$^{-1}$. 
For $T<358$~K, $\Delta Q$ is much smaller than the observed width of the 
PTCDA(102) reflection and cannot be resolved. 
For the narrow peaks observed at higher $T$, the peak splitting is not 
observed due to the dominance of the $\alpha$ phase at these temperatures.

Interestingly, the narrowing of the PTCDA(102) peak as a function of the 
growth temperature  reported here  has also been observed 
for PTCDA grown on H-passivated Si(100) \cite{Salvan00}, 
for PTCDA on Au(111) \cite{Fenter97}, 
and for annealed films of PTCDA \cite{Krausechemphys03}.  

\subsection{Modelling of the rod measurements}
\label{Modelling of the rod measurements}

Based on the morphological features observed with the AFM \cite{KrausePRB02}, 
the following model for the X-ray data has been developed. 

\begin{figure}[tbp!]
\centering
\includegraphics[width=0.85\columnwidth]{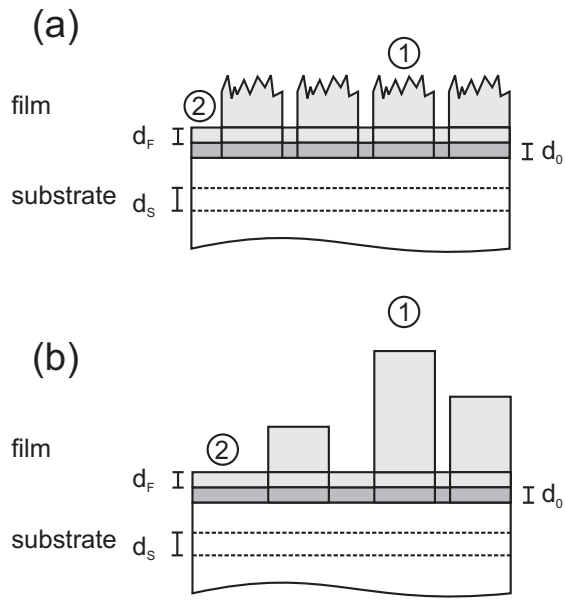}
\caption{Schematic of the morphology assumed for the simulations. 
(a) Low temperature model for $T \lesssim 358$~K, 
(b) high temperature model for $T \gtrsim 358$~K.}
\label{Fig8}
\end{figure}

In kinematic approximation the specular x-ray scattering intensity 
of each mosaic block of the substrate covered by a thin film is 
$I(q_z)=|\langle S(q_z) \rangle|^2$ with $S(q_z)$ as described in Eq. \ref{Eq1}. 
Since the morphology discussed in this section is observed long after 
the growth process and therefore time-independent, 
the parameter $t$ of  Eq. \ref{Eq1} has been omitted. 

Fig.\ \ref{Fig8} (a) shows a schematic of the model applied to the rod measurements 
with a temperature-independent peak width (comparatively smooth film). 
Two types of mosaic blocks have been introduced to describe this low-$T$ morphology: 
(1) mosaic blocks with $n_F$ film layers and a Gaussian roughness $\sigma_F$, 
described by
\begin{equation}
\label{Eq5}
\theta_k=0.5 \left[1- erf \left( \frac{k-n_F}{\sigma_F}\right)\right]
\end{equation}
and (2) mosaic blocks covered by a film of only a few monolayers and
\begin{equation}
\label{Eq6}
\theta_k = \bigg \{\begin{array}{l} \geq 0 \;\;\; \mbox{for} \;\;\;   k\leq 3 \\
        0 \;\;\; \mbox{for}  \;\;\;  k  > 3 
        \end{array}
\end{equation}
where $k$ is the layer index.
The total intensity scattered from the illuminated area is   
\begin{equation}
\label{Eq7}
I(q_z) = (1-A_2) \left| \langle S_1(q_z) \rangle \right|^2 +A_2\left|\langle S_2(q_z) \rangle \right| ^2, 
\end{equation}
where $S_1(q_z)$ and $S_2(q_z)$ are the scattering amplitudes of the different types 
of mosaic blocks, and $A_2$ is the fraction of the substrate surface covered 
by the type-(2) mosaic blocks. 

For the rod measurements with a decreasing peak width (islands on a wetting layer, high-$T$ films), the model is slightly modified (see Fig.\ \ref{Fig8} (b)). The islands (i.e.\ the PTCDA film of the type (1) mosaic blocks) are assumed to be molecularly flat with
\begin{equation}
\label{Eq8}
\theta_k = \bigg \{\begin{array}{l} 1 \;\;\; \mbox{for} \;\;\;   k\leq n \\
        0 \;\;\; \mbox{for}  \;\;\;  k  > n. 
        \end{array}
\end{equation}
They have a Gaussian height distribution with the average height $n_F$ 
and the standard deviation $\sigma_F$. The scattered intensity of the film is

\begin{eqnarray}
\label{Eq9}
\label{total intensity of the illuminated area high T}
\noindent
I(q_z) &=& (1-A_2) \frac{ \sum_{n} \left( e^{-\frac{\left(n-n_F\right)^2}{2 \sigma_F^2}}\left| \langle S_1(q_z,n) \rangle \right|^2 \right) }{\sum_{n} \left( e^{-\frac{\left(n-n_F\right)^2}{2 \sigma_F^2}} \right) } \nonumber \\
& &+A_2\left| \langle S_2(q_z) \rangle \right| ^2.
\end{eqnarray}
where $A_2$ is the area covered by the type-(2) mosaic blocks corresponding 
to the wetting layer, and $S_1(q_z,n)$ and  $S_2(q_z)$ are the scattering amplitudes 
of the different types of mosaic blocks. 

\begin{figure}[tbp!]
\centering
\includegraphics[width=0.85\columnwidth]{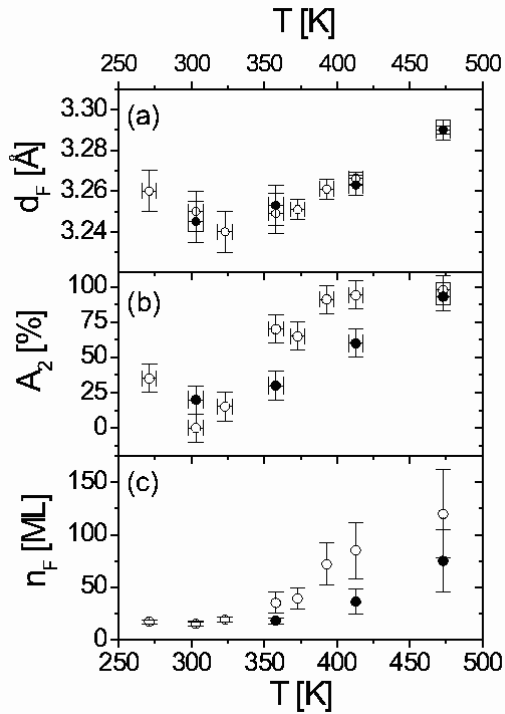}
\caption{Parameters used for the simulations shown in Fig.\  \ref{Fig1}. 
Open symbols: low $F$ data, filled symbols high $F$ data. 
(a) Lattice parameters of the PTCDA film as a function of $T$, 
(b) area covered by the type-(2) mosaic blocks corresponding to the wetting layer, 
and (c) height $n_F$ of the type-(1) mosaic blocks corresponding to areas 
covered by a thicker PTCDA film. 
The error bars of $n_F$ indicate $\sigma_F$, i.e.\ the roughness in the 
low temperature model and the island height distribution in the high temperature model, 
respectively.}
\label{Fig7}
\end{figure}
These two models describe all specular X-ray measurements. 
The results of the calculations are shown as straight lines in Fig.\ \ref{Fig1}. 
Fig.\ \ref{Fig7} summarizes the parameters used for the simulations. 
For $T \gtrsim 358$~K, the lattice parameter varies as expected from the 
thermal expansion of PTCDA in (102) direction \cite{Krausechemphys03}. 
For $T \lesssim 358$~K, the effective lattice parameter seems to be approximately 
temperature independent instead of decreasing further. 
A possible explanation for this is that the ratio of the PTCDA $\beta$ phase 
with a slightly larger lattice parameter than the $\alpha$ phase increases 
for $T \lesssim 358$~K as already indicated by in-plane measurements of 
thin PTCDA films \cite{KrausePRB02}.

The area covered by the wetting layer and the island height increase with $T$ for $T \gtrsim 358$~K, as observed in a similar AFM study \cite{KrausePRB02}. Since the island density increases with increasing deposition rate and decreasing growth temperature, the islands grown at high $F$ are always smaller than the islands grown at the same $T$ but lower $F$. 

The thickness of the wetting layer is approximately 2~ML and the lattice spacing between the topmost layer of the substrate and the first layer of the PTCDA film is  $d_0=2.8 \pm 0.1$~\AA, independent of $T$. 

For PTCDA on Au(111), similar measurements of the specular rod have been performed \cite{Fenter97}. These measurements have been modeled as the coherent scattering of a film with a laterally averaged density. In contrast to this, here the incoherent addition of the intensity has been chosen since the AFM images of PTCDA/Ag(111) show that the morphology varies on a length scale much larger than the average mosaic block size of 500~\AA, and the scattering of different mosaic blocks adds up incoherently.

\subsection{In situ growth experiments}
In the following, two aspects of the real-time growth observations will be discussed 
in detail, the asymptotic intensity, $I(q_z^\ast, \infty)$, and the intensity 
increase observed immediately after the deposition. 
Both features are related to the morphology of the films. 

\subsubsection{Asymptotic intensity}
\label{Asymptotic intensity}
For all samples, a constant asymptotic intensity $I(q_z^\ast, \infty)$ 
has been observed after the initial growth stages. 
With Eq.\ \ref{Eq3}, the experimentally observed value 
$I(q_z^\ast, \infty)=0.65 \pm 0.02$ corresponds to $\Delta \theta=0.5 \pm 0.03$.

In the following a model for this asymptotic intensity is developed. 
For simplicity only the  coherent scattering from one mosaic block is discussed. 
Strictly, this model only applies to the case where the roughness of the film 
is much smaller than the length scale of the mosaic block, 
as observed for the later growth stages of the relatively homogenous films 
grown at $T \lesssim 350$~K. 
At higher growth temperatures, the scattering from the PTCDA films after the 
initial growth stages is a superposition of the incoherent scattering of 
different types of mosaic blocks, as already discussed in 
Sec.\ref{Modelling of the rod measurements}. However, the incoherent addition 
does not change significantly the value for the asymptotic intensity, 
and the calculations can be performed in analogy to the coherent case.  

\begin{figure}[tbp!]
\centering
\includegraphics[width=0.85\columnwidth]{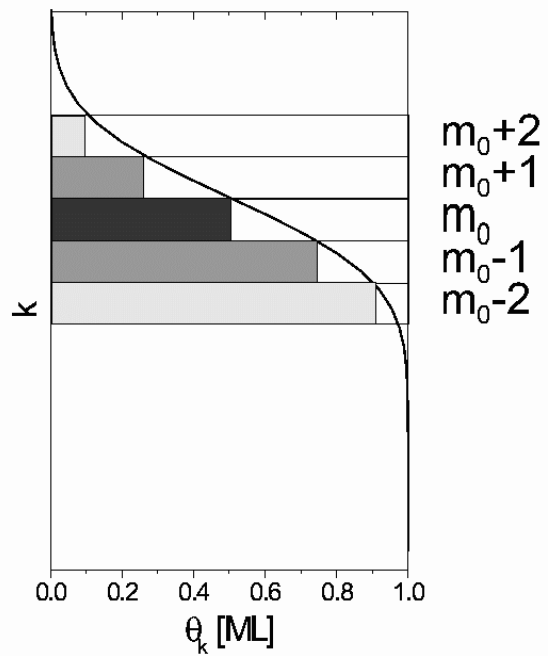}
\caption{Schematic explaining the asymptotic intensity. The roughness of the surface is assumed to be Gaussian. The average sample surface corresponds to the layer $m_0$. Partially filled layers $k$ in the same gray scale add up to a competely filled monolayer.}
\label{Fig5}
\end{figure}
For a Gaussian height distribution (which has been found on a length scale 
of the x-ray coherence length, approximately 1~$\mu$m, for films with 
$T \lesssim 350$~K after growth \cite{KrausePRB02}), 
the coverages $\theta_k$ of a film with an average layer number $m_0$ are determined by
(see Eq.\ \ref{Eq5})
\begin{equation}
\label{Eq10}
\theta_k=\frac{1}{2}\left(1-erf\left(\frac{k-m_0}{\sigma_F}\right)\right)
\end{equation}
(Fig. \ref{Fig5}). For sufficiently large $\sigma_F$, Eq.\ \ref{Eq10} can be characterized by
\begin{eqnarray}
\label{Eq11}
\theta_{m_0+\Delta} &=& 1-\theta_{m_0-\Delta}\nonumber\\
\mbox{and}\hspace{2cm}   \theta_{m_0} &=& 0.5.
\end{eqnarray} 
A function with these features leads to
$\Delta \theta\approx 0.5$ and thus $I(q_z^\ast, \infty)=0.65$,
which is shown in the Appendix.

For the system PTCDA/Ag(111), two cases have to be distinguished: 
The growth at $T \lesssim 350$~K, and the growth at $T \gtrsim 350$~K.

For $T \lesssim 350$~K, the layer-by-layer growth of the initial layers 
(indicated by the oscillations) changes with increasing deposition time to a 
Gaussian height profile explaining the asymptotic intensity during and the specular rod 
measurements after deposition (see Sec.\ \ref{Modelling of the rod measurements}). 
This case is consistent with the above model.

At $T \gtrsim 350$~K, separate islands on top of a wetting layer are observed after growth. 
This morphology deviates clearly from a symmetric roughness profile. 
The dominant part of the sample is covered by two monolayers ($\Delta \theta=0$), 
corresponding to $I(q_z^\ast,t)=1$. The islands give only a negligible contribution 
to the measured intensity since they cover only a small part of the area. 
In addition to this, only the topmost layer of the island contributes 
to the scattering since the scattering of all lower layers interferes destructively 
at the anti-Bragg point. However, one possible explanation for the observed intensity 
corresponding to $\Delta \theta=0.5$ is that during growth the third layer is partially 
filled with a coverage of approximately 0.5~ML. This coverage is not stable, 
and after growth, the molecules diffuse to the islands and only the 2 wetting monolayers 
observed in the rod measurements remain. In the next section, we will see that this model 
is consistent with the observation of the intensiy recovery after growth.

\subsubsection{Intensity recovery after deposition}
\label{Intensity recovery after deposition}
For samples deposited at growth conditions corresponding to islands on a wetting layer, 
a post-growth intensity recovery has been observed, 
while for samples grown at lower temperatures the intensity does not change after 
growth (on accessible time scales of several hours). 
During growth, the coverage changes of the different layers are due to deposition 
and diffusion. The post growth intensity changes are a sign for interlayer 
diffusion after growth 
(desorption is a negligible effect in the temperature range studied here).

The following model is consistent with the temperature dependence of the 
intensity recovery. For $T \lesssim 350$~K, after growth the samples have a 
Gaussian roughness on a scale much smaller than the mosaic block size. 
This roughness is not significantly changed by surface diffusion because 
$T$ is too low. Therefore, the scattered intensity does not change after growth. 
For $T \gtrsim 350$~K, post-growth diffusion of molecules from the top 
of the wetting layer to the islands takes place.

\begin{figure}[tbp!]
\centering
\includegraphics[width=0.85\columnwidth]{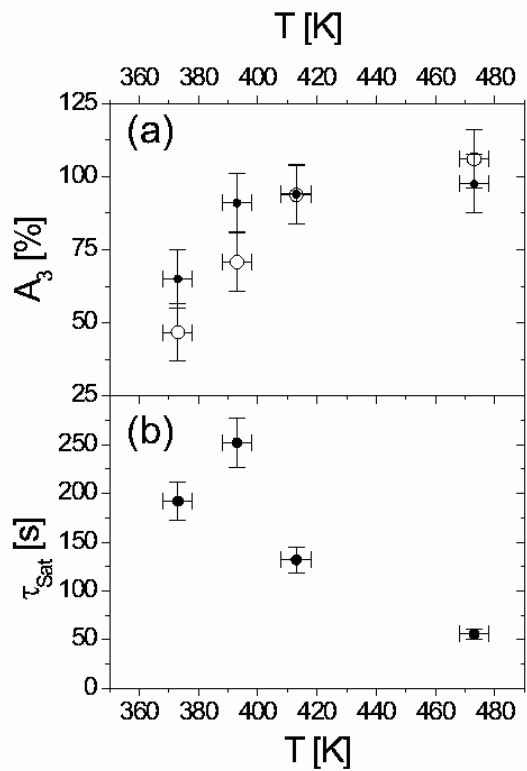}
\caption{(a) Area covered by the wetting layer as a function of $T$. Open symbols: fit result of the post growth intensity changes; filled symbols: result of the fit of the specular rod measurements. (b) Characteristic time for the post growth intensity changes as a function of $T$.}
\label{Fig6}
\end{figure}
To model the x-ray scattering of the inhomogeous morphology at $T \gtrsim 350$~K, the scattered intensity of 3 types of mosaic blocks is added incoherently: (1) Mosaic blocks belonging to islands with even layer number, the coverage difference $\Delta \theta=0$, and the normalized scattering intensity $I_{even}=1$ according to Eq.\ \ref{Eq3}; (2) mosaic blocks belonging to islands with odd layer number, the coverage difference $\Delta \theta=1$, and the normalized scattering intensity $I_{odd}=1+A+B$; (3) mosaic blocks covered by the wetting layer and some additional molecules, with the time-dependent scattering intensity $I_3(t)=1+A \Delta \theta + B (\Delta \theta)^2$ and the coverage difference $\Delta \theta_{ML}(t)= 0.5 e^{-t/\tau_{Sat}}$ covering the area $A_3$. $\tau_{Sat}$ is an effective parameter comprising the surface diffusion barrier, the size of the critical nucleus, and the temperature dependence of the surface diffusion. We assume that the islands with even and odd layers each cover half of the remaining area, which is $0.5(1-A_3)$. The total scattered intensity is
\begin{equation}
\label{Eq16}
I(t)=A_3 I_3(t) + 0.5(1-A_3)( I_{even}+I_{odd}).
\end{equation}
Fig.\ \ref{Fig6} shows the fit results of the intensity measurements after deposition. 
The area covered (only) by the wetting layer (Fig.\ \ref{Fig6} (a)) 
increases with the deposition temperature up to nearly the entire sample surface. 
This is consistent with the result from specular rod measurements, 
also shown in Fig.\ \ref{Fig6} (a), and the AFM observations \cite{KrausePRB02}. 
The characteristic time, $\tau_{Sat}$, for the recovery process is shown in 
Fig.\ \ref{Fig6} (b). 
It decreases with increasing $T$, as expected for a diffusion-related process, 
from $\tau_{Sat} \approx 200$~s 
at $T=373$~K to $\tau_{Sat} \approx 50$~s at $T=473$~K.

Similar to the x-ray intensity recovery after growth reported here, 
an intensity recovery after the growth of GaAs has been observed by 
Reflection High Energy Electron Scattering \cite{Lewis85,Joyce90}. 
In this case, the recovery corresponds to a smoothening of the surface 
after the interruption of the Ga flux due to the post-growth surface diffusion. 
In contrast to this, the post-growth intensity recovery of PTCDA is explained 
as the result of the interplay between post-growth surface diffusion processes 
and the morphology of the PTCDA films.

\section{Summary and Conclusions}
\label{Summary and Conclusions}

The later growth stages, the post growth diffusion, 
and the structure of thin films of PTCDA/Ag(111) have been studied. 
The specular rod data are strongly influenced by the temperature-dependent 
morphology of the films.

The thickness of the wetting layer (2 ML), the area covered by the islands, 
and the binding distance between the first monolayer and the substrate 
have been determined. 
{\sl In-situ} measurements at the anti-Bragg point have been used to derive 
information about the growth of PTCDA. 
After the intensity oscillations during the deposition of the wetting layer 
(indicating layer-by-layer growth), a constant intensity has been observed, 
followed by an increase of intensity after growth for $T \gtrsim 350$~K. 
The measurements have been analyzed quantitatively using the following growth model: 
At $T \lesssim 350$~K, after the deposition of the wetting layer the growth mode 
changes to 3D growth with a Gaussian roughness which is stable after growth. 
At $T \gtrsim 350$~K, the growth of the wetting layer is followed by island growth. 
During growth, the third layer is partially filled. 
After growth, due to the strong surface diffusion this material is also incorporated 
into the islands,
and the temperature dependence of the time scales of the post-growth diffusion
has been determined.

Morphology changes as a function of the growth temperature have a strong impact 
on the performance of optoelectronic devices. The observed post-growth diffusion 
may be of special interest for multilayer structures since the interface quality 
can be improved by growth interruptions allowing for the incorporation of residual 
surface molecules. 
Also, the resulting island morphology might be exploited similar as in the 
Stranski-Krastanov growth of inorganic semiconductors.

\begin{acknowledgments}
We are grateful to N.~Karl for providing the purified material. 
Partial support from the DFG (Schwerpunktprogramm ``Organische Feldeffekt-Transistoren'') 
is gratefully acknowledged. 
We thank the HASYLAB staff for excellent support.
\end{acknowledgments}
%

\appendix

\section{Asymptotic intensity}

It is shown that a function $\theta_k$ with the features described in Eq.\ \ref{Eq11} 
leads to $\Delta \theta\approx 0.5$ and thus $I(q_z^\ast, \infty)=0.65$. 
$\Delta \theta$ can be written as 
\begin{eqnarray}
\label{Eq12}
\Delta \theta &=& \sum_{k=1}^{\infty} \theta_k(-1)^{k-1} \nonumber \\
&=&\sum_{k=1}^{m_0-1} \theta_k(-1)^{k-1} + \sum_{k=m_0+1}^{\infty} \theta_k(-1)^{k-1} \nonumber \\
& & + \theta_{m_0}(-1)^{m_0-1}.
\end{eqnarray}
Here, the sum has been separated into the contribution of the layers with $k<m_0$, 
the layers with $k>m_0$, and the contribution of the layer $m_0$. 
With $k = m_0-\Delta$, we get
\begin{eqnarray}
\label{Eq13}
\Delta \theta &=& \sum_{\Delta=1}^{m_0-1}\theta_{m_0-\Delta}(-1)^{m_0-\Delta-1} \nonumber \\ 
& + & \sum_{\Delta=1}^{\infty}\theta_{m_0+\Delta}(-1)^{m_0+\Delta-1}+\theta_{m_0}(-1)^{m_0-1}.
\end{eqnarray}
With Eq.\ \ref{Eq11} and $(-1)^{m_0-\Delta}=(-1)^{m_0+\Delta}$ it follows
\begin{eqnarray}
\label{Eq14}
\Delta \theta &=& \sum_{\Delta=1}^{m_0-1}(1-\theta_{m_0+\Delta})(-1)^{m_0+\Delta-1} \nonumber \\
&+&\sum_{\Delta=1}^{\infty}\theta_{m_0+\Delta}(-1)^{m_0+\Delta-1}+\theta_{m_0}(-1)^{m_0-1} \nonumber \\
&=& \sum_{\Delta=1}^{m_0-1}(-1)^{m_0+\Delta-1} \nonumber \\
&+&\sum_{\Delta=m_0+1}^{\infty}\theta_{m_0+\Delta}(-1)^{m_0+\Delta-1}+\theta_{m_0}(-1)^{m_0-1}.
\end{eqnarray}
For $m_0 \gg \sigma_{PTCDA}$, and with $\theta_{m_0}=0.5$ this reduces to
\begin{equation}
\label{Eq15}
\Delta \theta=(-1)^{m_0-1}\left(\sum_{\Delta=1}^{m_0-1}(-1)^{\Delta}+0.5 \right)=0.5
\end{equation}
for both $m_0$ even and $m_0$ odd.

The physical reason for this is the following: 
If enough layers contribute to the Gaussian surface roughness, 
a layer $m_0$ can always be chosen with $\theta_{m_0} \approx 0.5$, 
as indicated in Fig.\ \ref{Fig5}. 
The X-rays scattered by the layer $m_0+1$ interfere constructively 
with the X-rays scattered by the layer $m_0-1$. 

If the coverage of the layer $m_0+1$ is shifted to the layer $m_0-1$, 
the specularly scattered intensity at the anti-Bragg point does not change. 
Because of the symmetry of the Gaussian lineshape, 
the coverage of the layer $m_0+1$ fills exactly the vacancies of the layer $m_0-1$, 
and all layers $k<m_0$ can be filled with the coverage of a corresponding layer 
with $m_0<k<2m_0$. If $m_0$ is large enough compared to the roughness of the film, 
the layers with $k\geq 2m_0$ have the coverage 0. 
The total specular intensity at the anti-Bragg point scattered by the rough film 
is equivalent to the intensity scattered by $m_0-1$ layers with the coverage 1 
(corresponding to Eq.\ \ref{Eq15}), and by the layer $m_0$ with the coverage 0.5. 
Therefore, $\Delta \theta$ always equals 0.5.


\end{document}